\documentclass[submission, Proceedings]{SciPost}

\hypersetup{
    colorlinks,
    linkcolor={red!50!black},
    citecolor={blue!50!black},
    urlcolor={blue!80!black}
}

\usepackage[bitstream-charter]{mathdesign}
\DeclareSymbolFont{usualmathcal}{OMS}{cmsy}{m}{n}
\DeclareSymbolFontAlphabet{\mathcal}{usualmathcal}

\usepackage{xspace}      
\newcommand{\epem}{\ensuremath{\text{e}^+\text{e}^-}\xspace}
\newcommand{\textdef}[1]{\emph{#1}} 
\newcommand{\ttbar}{\ensuremath{\text{t}\bar{\text{t}}}\xspace}

\begin{document}

\begin{center}{\Large \textbf{
Top-quark mass determination in the optimised threshold scan 
}}\end{center}

\begin{center}
Kacper Nowak\textsuperscript{$\star$} 
and Aleksander Filip \.Zarnecki
\end{center}

\begin{center}
Faculty of Physics, University of Warsaw, Poland
\\
* k.nowak27@student.uw.edu.pl
\end{center}

\begin{center}
\today  
\end{center}


\definecolor{palegray}{gray}{0.95}
\begin{center}
\colorbox{palegray}{
  \begin{tabular}{rr}
  \begin{minipage}{0.1\textwidth}
    \includegraphics[width=22mm]{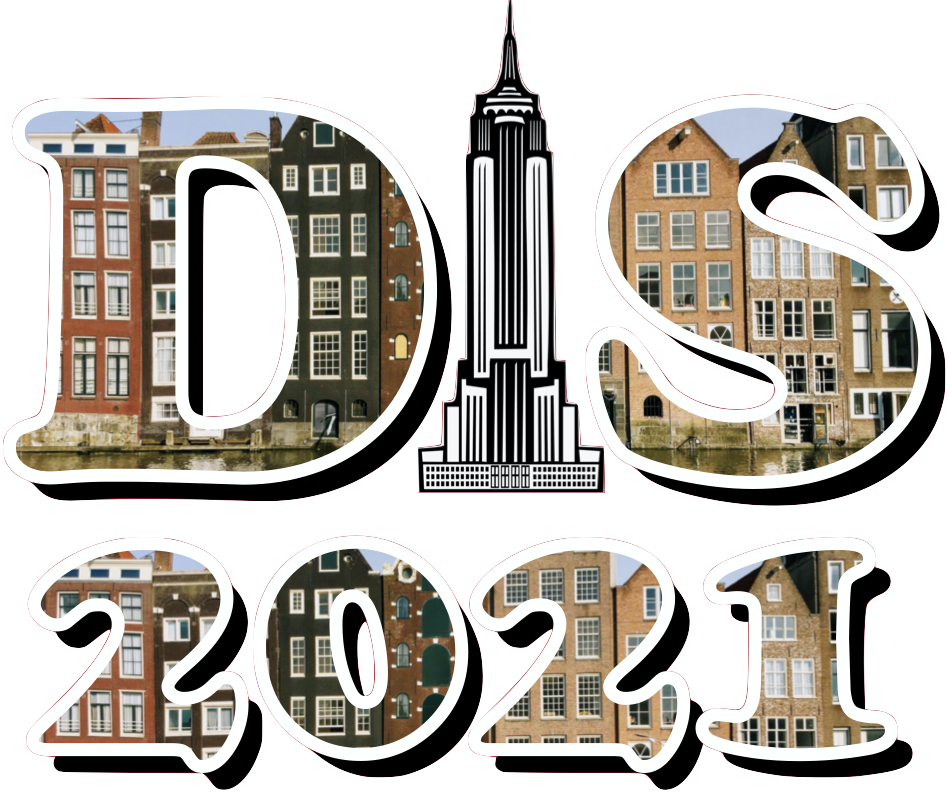}
  \end{minipage}
  &
  \begin{minipage}{0.75\textwidth}
    \begin{center}
    {\it Proceedings for the XXVIII International Workshop\\ on Deep-Inelastic Scattering and
Related Subjects,}\\
    {\it Stony Brook University, New York, USA, 12-16 April 2021} \\
    \doi{10.21468/SciPostPhysProc.?}\\
    \end{center}
  \end{minipage}
\end{tabular}
}
\end{center}

\section*{Abstract}
{\bf
One of the important goals at the future e$^+$e$^-$ colliders is to
measure the top-quark mass and width in a scan of the pair production
threshold.  Presented in this work is the most general approach to
the top-quark mass determination from the threshold scan at CLIC, with
all relevant model parameters and selected systematic uncertainties
included in the fit procedure. In the baseline scan scenario the top-quark mass can be
extracted with precision of the order of 30 to 40\,MeV, already for 100 fb$^{-1}$ of data
collected at the threshold. We present the optimisation procedure based
on the genetic algorithm with which the statistical uncertainty of the mass
measurement can be reduced by about 20\%. 
}

\section{Introduction}
\label{sec:intro}

The threshold scan is expected to be the most precise way to measure the top-quark mass.
However, the threshold cross-section shape depends not only on the top-quark mass and width 
but also on other model parameters, such as the top Yukawa coupling and the strong coupling
constant. 
It can also be affected by many systematic effects.
Dedicated fit procedures have been developed for the top-quark
threshold scan analysis at CLIC \cite{wilga2019,Nowak:2021xmp}. The new fit procedure
is more flexible that the one used in the previous study
\cite{clic-top} and allows to include all relevant model parameters as
well as additional constraints on model parameters, coming e.g. from
earlier measurements,  
and constraints on data normalisation.

At the first energy stage, CLIC running is assumed to include a dedicated scan of the \ttbar threshold 
with total integrated luminosity of 100 fb$^{-1}$. The baseline scenario of the threshold scan
assumes running at 10 equidistant energy points taking 10 fb$^{-1}$ of
data for each value of collision energy.  
This running scenario seems to be conservative and the aim of the presented study was to investigate
to what extent statistical uncertainties can be reduced when using the optimised running scenario.
The scan optimisation assumes that the top-quark mass is already known to O(100\,MeV).

\section{Genetic algorithm}

When looking for the best scenario of the top-quark threshold
scan at CLIC (or at any other future \epem collider) one needs to
take into account many different aspects of the measurement.  
The top quark mass is just one of the parameters that are to be constrained from the collected data
(with the best possible statistical uncertainty). There are other 
model parameters, measurement of which needs to be optimised at the same time.
In a Genetic algorithm, a set of proposed solutions to an optimisation problem, 
called \textdef{Individuals}, is evolved towards better solutions. 
Each Individual has a defined set of properties, called
\textdef{genotype},  which can be mutated and altered, and a set of
measurable traits, called \textdef{phenotype}, that are determined by
the genotype.
Phenotype consists of traits that are used to evaluate their
performance and choose the best Individuals to the next generation. 
During consecutive iterations of the algorithm, population evolves
towards better solutions. 
Ultimately, after a finite number of iterations, the population should
converge to an optimal solution. We decided to use the Non dominated
sorting genetic algorithm II, proposed by Kalyanmoy Deb in 2002.
Thanks to the use of an efficient non dominated sorting algorithm it was possible to
lower the time complexity from  original $O(MN^2)$ \cite{deb} to $O(MN
log^{M-1}N)$ \cite{jensen}, where $M$ is number of objectives and $N$
is the size of the population. For more details refer to \cite{Nowak:2021xmp}.

\section{NSGA-II set-up}
One scan scenario was assumed as an Individual, which genotype is
represented by a scan sequence, set of centre-of-mass energy points. 
Each scan point can be considered a chromosome.
Constant total integrated luminosity of 100\,fb$^{-1}$ was assumed
and it is equally distributed among all scan points.
An initial population was created from the baseline scenario assuming ten scan points
equally separated from each other by 1\,GeV, starting at 340\,GeV. 
This scenario was studied in detailed in \cite{Nowak:2021xmp,clic-top}.
The population size is set to 2000 and number of generations to 30. All
results presented here were calculated assuming
normalisation uncertainty of $\Delta=0.1\%$, strong coupling constant
uncertainty of $\sigma_{\alpha_s}=0.001$ and background level
uncertainty of $\sigma_{f_{bg}}=2\%$.
When measurement of the top Yukawa coupling was not included in the
optimisation objectives, an uncertainty of $\sigma_{y_t} = 0.1$ was assumed.
In order to rule out results that could be not representative of Individual's 
performance, each uncertainty was computed three times and the worst 
result was chosen.

\section{Optimisation}
Described in this contribution are the results of the two objective optimisation procedures, 
optimised for top-quark mass and width or  mass and Yukawa coupling measurements. 
For results of single objective optimisation please refer to \cite{Nowak:2021xmp}.

    \begin{figure}[t]
        \centering
\includegraphics[height=5.5cm]{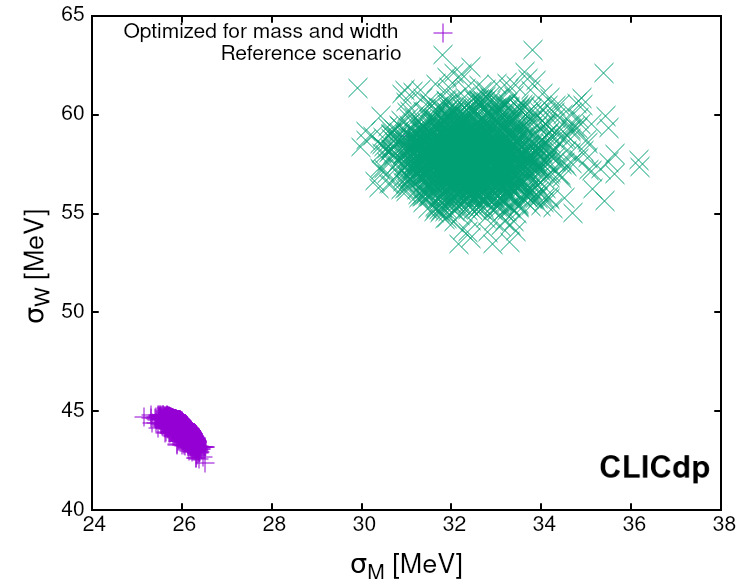}
\hspace{0.5cm}        
\includegraphics[height=5.5cm]{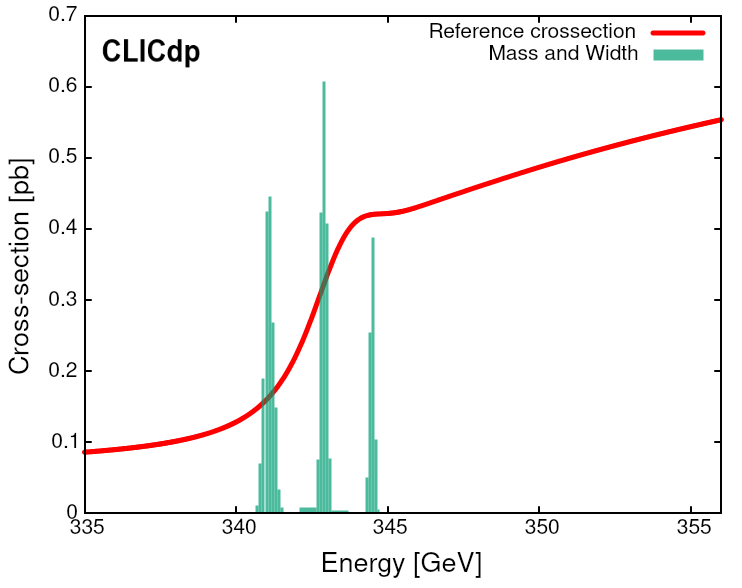}
        \caption{Left: mass and width uncertainty distribution in the first (green) 
        and the last (blue) generation for scan optimised for mass and
        width determination. Right: distribution of the measurement
        points from the last generation (arbitrary scale) compared
        with the reference cross section template. } 
        \label{fig:multiMW}
    \end{figure}
    \begin{figure}[t]
        \centering
\includegraphics[height=5.5cm]{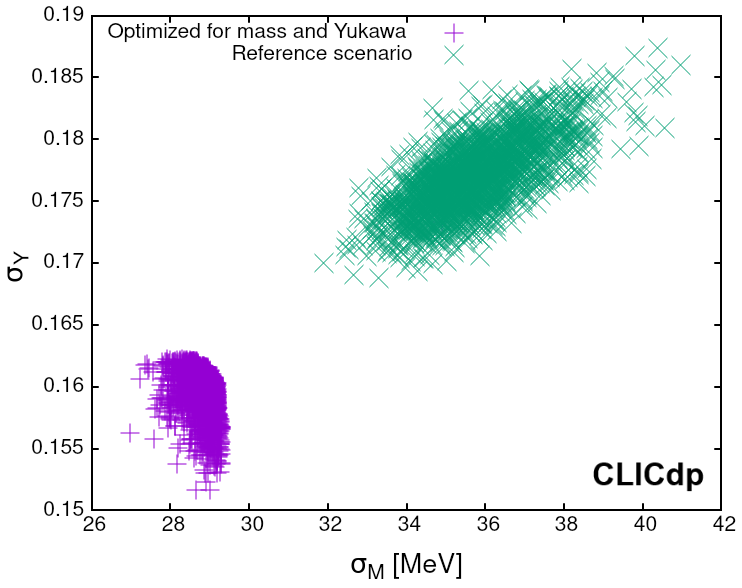}
\hspace{0.5cm}        
\includegraphics[height=5.5cm]{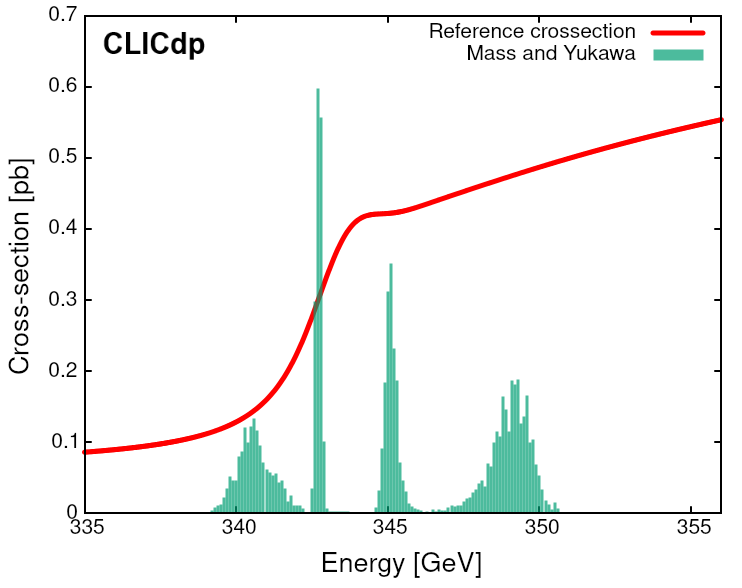}
        \caption{As in Fig.~\ref{fig:multiMW} but for scan optimised
          for mass and Yukawa coupling determination. } 
        \label{fig:multiMY}
    \end{figure}

When considering scan optimisation for the mass-width pair, an
improvement of around 20-25\% could be observed in the final
generation for both parameters.  
For the top-quark mass the uncertainty was reduced from 32\,MeV to 26\,MeV, 
while for the width, from 58\,MeV to 44\,MeV (see Fig \ref{fig:multiMW}).
If we look at the histogram of measurement points from the last
generation  (see Fig \ref{fig:multiMW}) we can clearly observe 
three distinct regions on the threshold. One below, one in the middle
and one above the threshold. We conclude that cross section
measurements in those regions are most sensitive to the two considered
model parameters, mass and width.  

When optimizing the threshold scan procedure for mass and Yukawa coupling measurements, 
we can observe a similar behaviour. 
The mass uncertainty changes from 32\,MeV to 28\,MeV, while the Yukawa
coupling uncertainty changes from 0.18 to 0.16. 
The very small improvement for the Yukawa coupling can be explained,
when we look at Fig. \ref{fig:multiMY}.  
On the left plot we can clearly see an additional region above the threshold, which is responsible for 
the determination of the Yukawa coupling. The benchmark scenario from
the initial generation has half of its points in this region, not
making significant improvement possible.  

\section{Optimised scan scenarios}

\begin{figure}[t]
        \centering
\includegraphics[height=5.5cm]{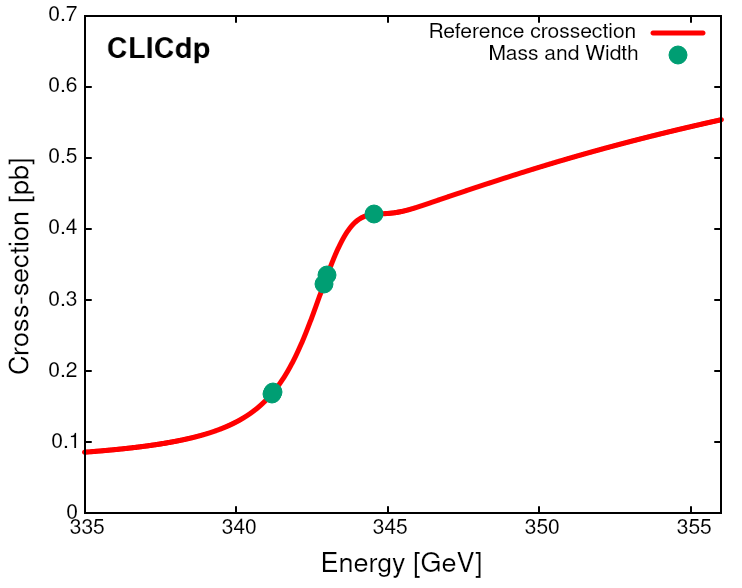}
\hspace{0.5cm}
\includegraphics[height=5.5cm]{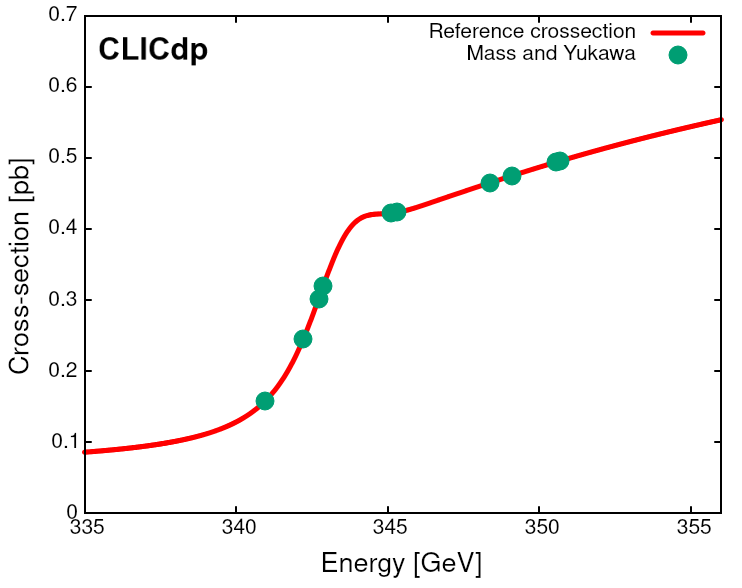}
        \caption{Scan energy points for the ``best scenarios'' taken from the last generation compared with 
        the reference cross section template: (left) 5 point scenario optimised for mass and width determination 
        precision (two points below, two in the middle and one above the threshold) and (right) 
        10 point scenario optimised for mass and Yukawa coupling determination precision. }
        \label{fig:scen}
    \end{figure}

To verify the optimisation results we select one test scenario from the final generation
for each of the two considered optimisation configurations (See Fig. \ref{fig:scen}).
The algorithm converged well in both cases. For mass and width optimisation 99\% of final generation
scenarios had 5 scan points, while for mass and Yukawa 97\% of them had 9 or 10 scan points.
To confirm optimisation results we generated 20\,000
pseudo-experiments using selected running scenarios.  

\begin{figure}[t]
\centering
\includegraphics[height=5.5cm]{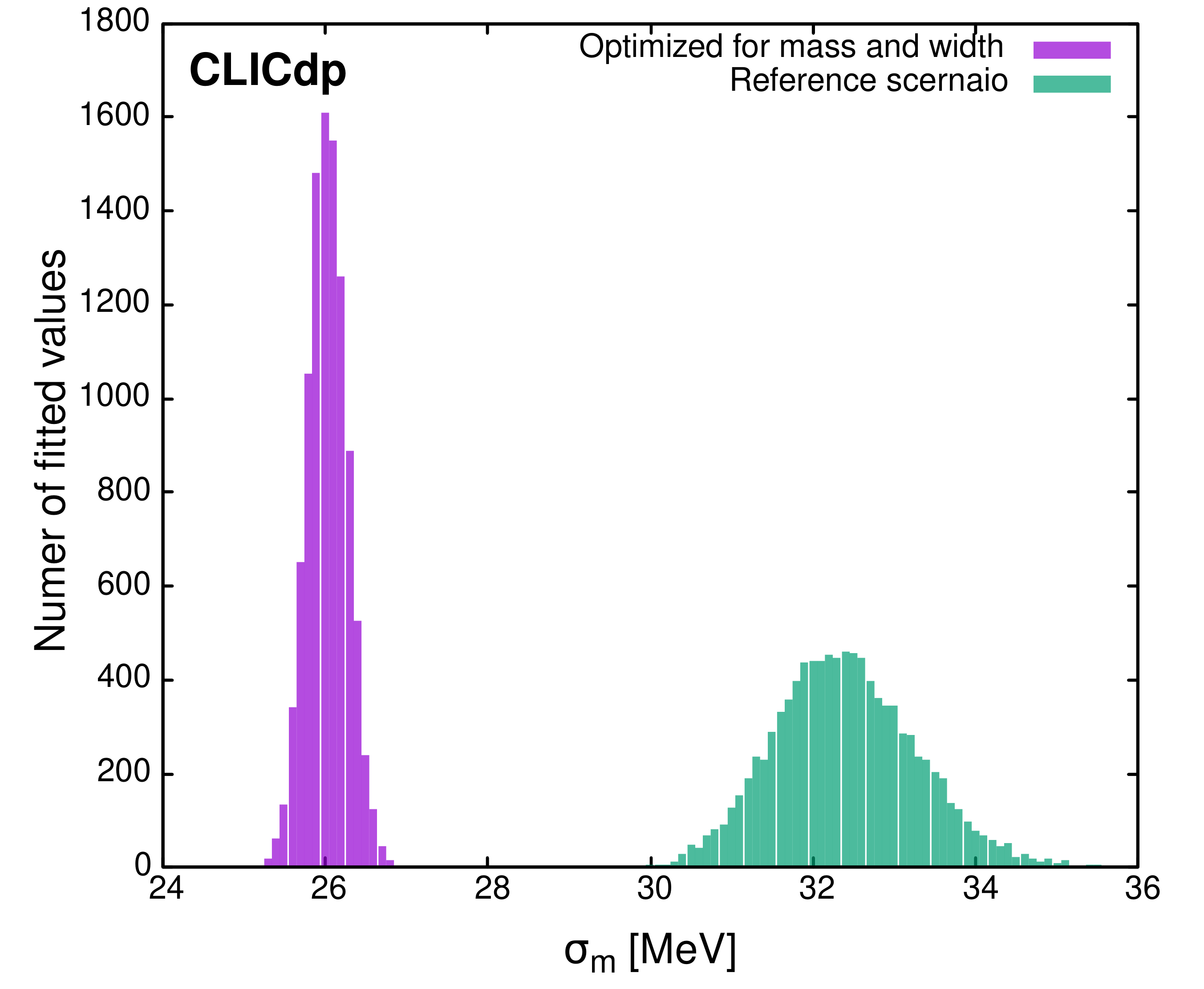}
\hspace{0.5cm}
\includegraphics[height=5.5cm]{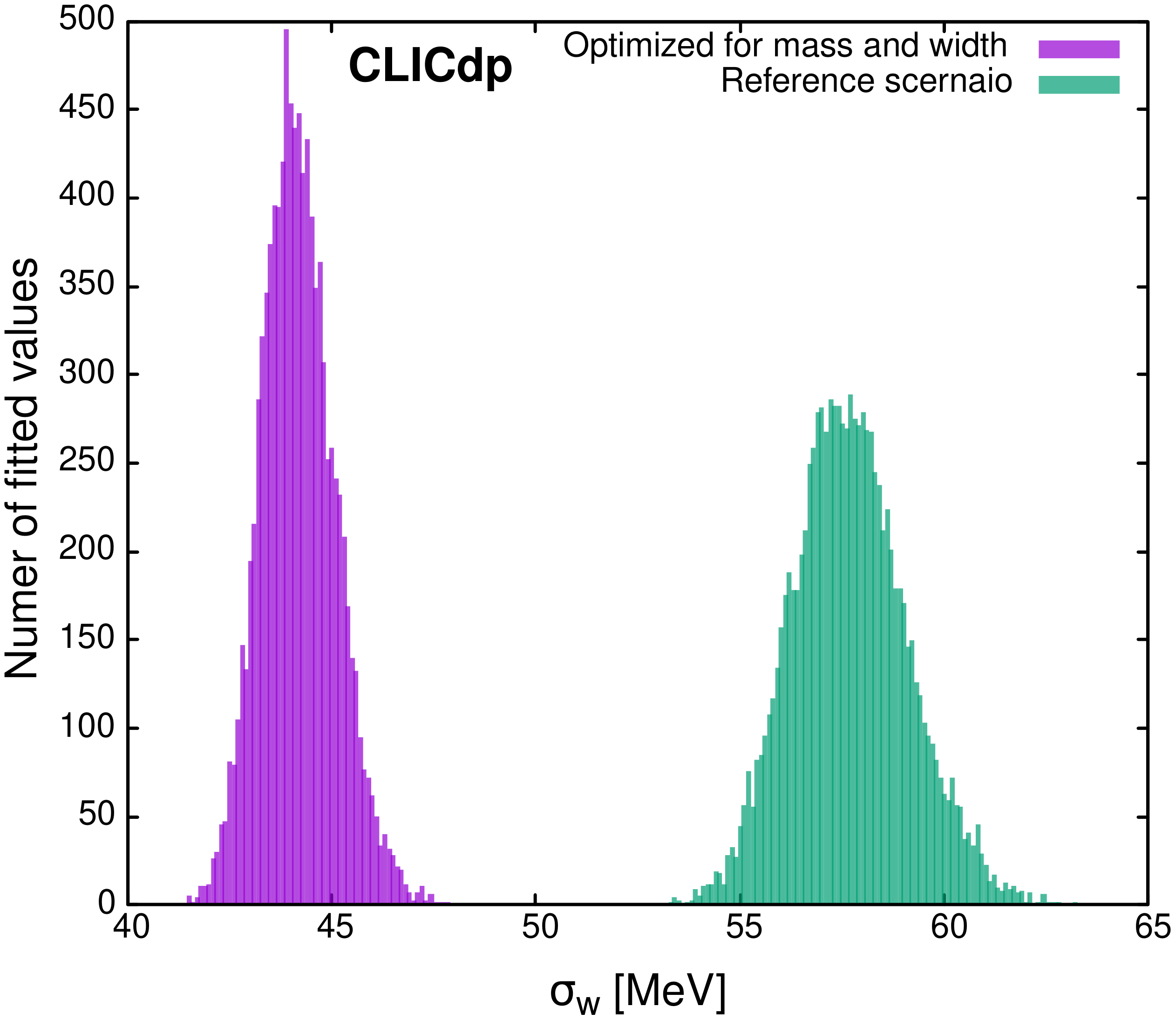}
        \caption{Uncertainty distribution for mass (left) and width (right) measurement, 
        for five-point scan scenario optimised for mass and width measurement 
        (see Fig. \ref{fig:scen}), compared with the distributions for the reference scenario.  }
        \label{fig:histMW}
    \end{figure}
    
 \begin{figure}[tb]
        \centering
\includegraphics[height=5.5cm]{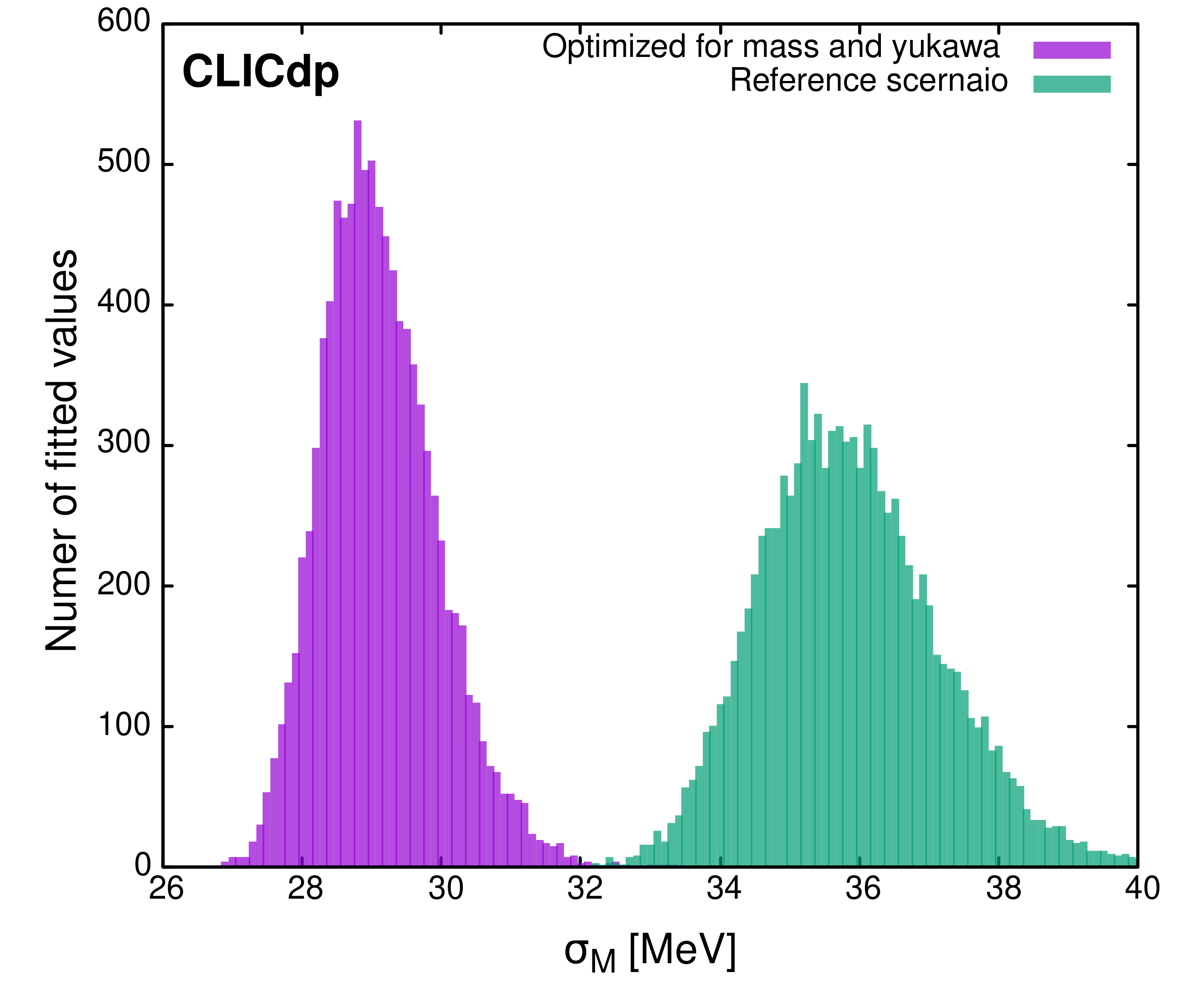}
\hspace{0.5cm}        
\includegraphics[height=5.5cm]{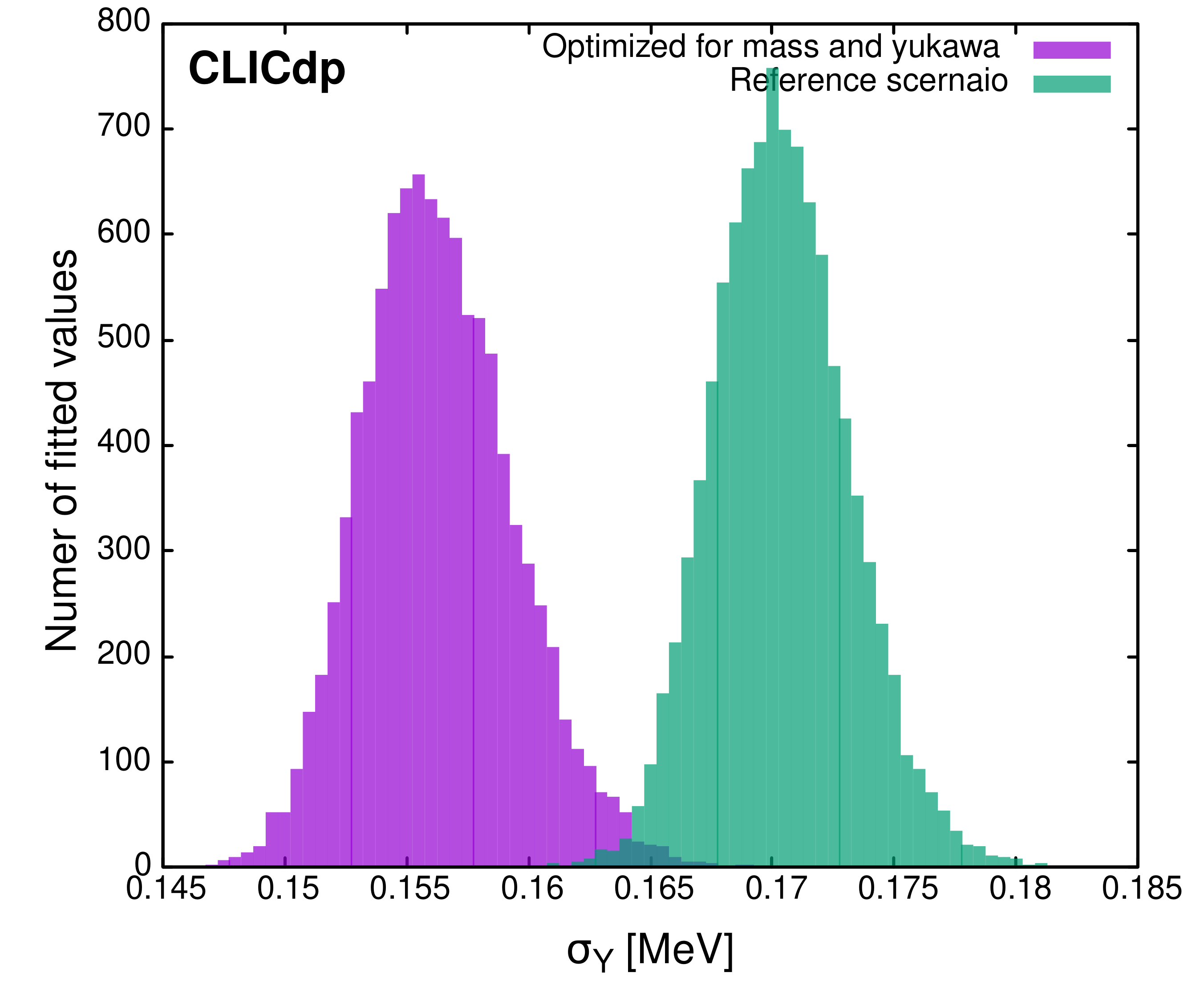}
      \caption{Uncertainty distribution for mass (left) and Yukawa coupling (right) measurement, 
        for 8 point scan scenario optimised for mass and Yukawa coupling measurement 
        (see Fig. \ref{fig:scen}), compared with the distributions for the reference scenario.  }
        \label{fig:histMY}
    \end{figure}
    
Results obtained with a large namber of pseudo-experiments confirm
estimates from the optimisation procedure (where each
scenario was evaluated based on three pseudo-data sets only).
For the optimised mass-width scenario the average expected mass uncertainty 
is around 26\,MeV, while for width it is around 44\,MeV, see Fig.~\ref{fig:histMW}.
Furthermore, uncertainty distributions are narrower than those for the
reference scenario confirming that the fit is very stable and less
sensitive to the  statistical  fluctuations, which is the result
of including three pseudo-experiments for each Individual. 

For scenario optimised for mass and Yukawa coupling measurement, 
results obtained from 20 000 of pseudo-experiments, shown in
Fig.~\ref{fig:histMY} are again in good agreement with optimisation
results (see Fig.~\ref{fig:multiMY}).
However, uncertainty distribution for the mass
measurement is significantly wider than the one obtained for mass and
width optimised scenario (Fig.~\ref{fig:histMW}).
It is also slightly asymmetric, with a larger tail towards high
uncertainty values. Nevertheless, the optimised scenario provides better
mass measurement precision than the reference one in every case.

\section{Conclusions}
An optimisation procedure using a non-dominated sorting genetic
algorithm II has been applied to the top-quark pair-production threshold scan.
Each measurement scenario (set of energy points with total equally
distributed luminosity of 100 fb\,$^{-1}$) is considered a genotype
and results of the fit procedure constitute a phenotype.
Starting from the benchmark scenario (10 scan points, equally separated from each other),
stable optimisation results are obtained from the genetic evolution 
for 30 generations and population size of 2000.
By using this optimisation procedure, it was shown that
it is possible to reduce the top-quark mass uncertainty by up to 20\%, without
increase of luminosity or loss of precision in determination of other parameters. 

\section*{Acknowledgements}

The work was carried out in the framework of the CLIC detector and
physics (CLICdp) collaboration.
We thank collaboration members for fruitful discussions, valuable comments and suggestions.
The work was partially supported by the National Science Centre
(Poland) under OPUS research projects no. 2017/25/B/ST2/00496
(2018-2021).

\bibliography{dis2021_top.bib}

\nolinenumbers

\end{document}